\documentclass[twocolumn]{aastex62}

\graphicspath{{./}{figures/}}

\received{ }
\revised{ }
\accepted{ }

\submitjournal{ApJ}

\shorttitle{H Masers and Dasars  in Mz3}
\shortauthors{Abraham et al.}
\begin{document}

\title{ALMA detection of Masers and Dasars in the Hydrogen Recombination Lines of the Planetary Nebula Mz3}

\correspondingauthor{Zulema Abraham}
\email{zulema.abraham@iag.usp.br}

\author[0000-0003-0012-1315]{Zulema Abraham}
\affil{Instituto de Astronomia, Geof\'isica e Ci\^encias Atmosf\'ericas, Universidade de S\~ao Paulo \\
Rua do Mat\~ao 1226, CEP 05508-090, S\~ao Paulo, Brazil\\}

\author{Pedro P. B. Beaklini}
\affiliation{National Radio Astronomy Observatory, 1003 Lopezville Road, Socorro, NM 87801, USA}

\author{Isabel Aleman}
\affiliation{Institute of Mathematics and Statistics, University of S\~{a}o Paulo, Rua do Mat\~{a}o, 1010 - Butant\~{a}, 05508-090, S\~{a}o Paulo, SP, Brazil\\}
\affiliation{Universidade Federal de Itajub\'a, Instituto de F\'isica e Qu\'imica, Av. BPS 1303 Pinheirinho, 37500-903 Itajub\'{a}, MG, Brazil\\
}
\author{Raghvendra Sahai}
\affiliation{Jet Propulsion Laboratory, MS 183-900, California Institute of Technology, Pasadena, CA 91109, USA}

\author{Albert Zijlstra}
\affiliation{Jodrell Bank Centre for Astrophysics, Alan Turing Building, University of Manchester, Manchester, M13 9PL, UK}
\author{Stavros Akras}
\affiliation{Institute for Astronomy, Astrophysics, Space Application \& Remote Sensing, National Observatory Athens, GR-15236, Athens, Greece}

\author{ Denise R. Gon\c calves}
\affiliation{Valongo Observatory, Federal University of Rio de Janeiro, Ladeira Pedro Antonio 43, 20080-090 RJ, Brazil
}
\affiliation{Department of Space, Earth and Environment, Chalmers University of Technology,
SE-412 96, Gothenburg, Sweden}

\author{ Toshiya Ueta}
\affiliation{Dept. of Physics and Astronomy, University of Denver, 2112 E Wesley Ave., Denver, CO 80210}

\begin{abstract}

The hydrogen recombination lines  H30$\alpha$, H40$\alpha$, H42$\alpha$, H50$\beta$ and H57$\gamma$ and the underlying bremsstrahlung continuum emission were detected with ALMA in the bipolar nebula Mz3.  
The source was not spatially resolved, but the velocity profile of the H30$\alpha$ line shows clear indication of maser amplification, confirming previous reports of laser amplification in the  far infrared H recombination lines  observed with Herschel Space Observatory. 
Comparison between the flux densities of the H50$\beta$, H40$\alpha$ and H42$\alpha$  lines show overcooling, or darkness amplification by stimulated absorption  (dasar effect) at the LSR velocity of about $-25$ km s$^{-1}$, which constrains the density of the  absorbing region to about 10$^3$ cm$^{-3}$. The H30$\alpha$ line, on the other hand, presents maser lines at LSR velocities of $-69$  and $-98$ km s$^{-1}$, which indicates ionized gas with densities close to 10$^7$  cm$^{-3}$. 
Although the source of emission was not resolved, it was possible to find the central position of the images for each  velocity interval, which resulted in a well defined position-velocity distribution. 
\end{abstract}
\section{Introduction}
\label{introduction}
Menzel 3 (Mz3; Ant Nebula; PN G331.7$-$01.0),  is seen as a bipolar nebula in  optical light, with a high degree of axial symmetry. It exhibits pairs of multiple nested outflows, an also bipolar structure of spokes, and a puzzling ring structure surrounding its waist in a tilted position with respect to the main nebular symmetry axis \citep["chakram";][]{Guerrero_etal_2004,SantanderGarcia_etal_2004, clyne_etal_2015}. 
\begin{figure}
\includegraphics[width=8.5cm]{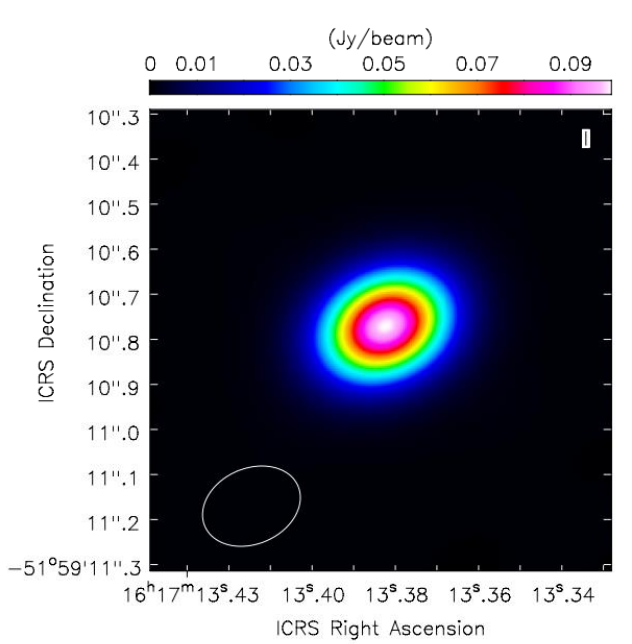}
    \caption{Continuum image of Mz3 obtained with ALMA at 93 GHz. The color wedge is presented at the top of the graph and the beam ellipse at the bottom left corner.}
    \label{fig_1}
    \end{figure}
\begin{figure*}
	\includegraphics[width=18 cm]{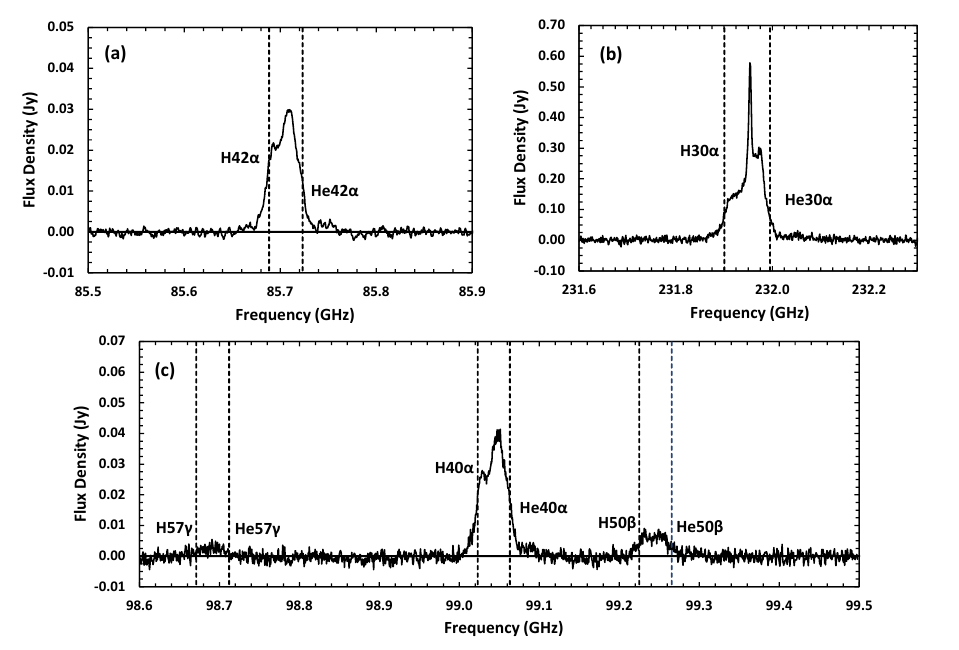}
    \caption{(a) H42$\alpha$, (b) H30$\alpha$, (c)  H40$\alpha$, H$50 \beta$ and H57$\gamma$ line profiles of Mz3. The vertical lines represent the rest frequencies of the respective H and He lines.  Notice the factor of ten  difference in scale between the H30$\alpha$  and the other two lines
    }
    \label{fig_2}
\end{figure*}
\begin{figure}
	\includegraphics[width=8.5 cm]{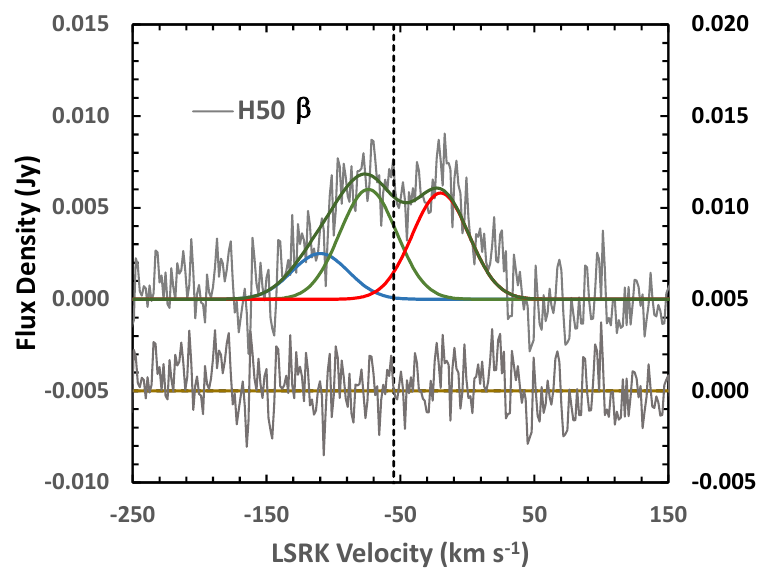}
    \caption{Observed H50$\beta$ line (gray line), together with a 3-gaussian fit (individual gaussians are shown in blue, green and red, the sum is shown in black); the flux density of the lines (top plot) is shown in the left vertical axis, and the residuals of the fitting (bottom plot) in the right vertical axis.  the vertical dashed line represents the systemic velocity of Mz3
    .
    The parameters of the gaussians  (amplitude in Jy, center velocity  and half power width in km s$^{-1}$) are: $(0.0025, -110, 20.8)$, $(0.0060, -74, 20.8)$ and $(0.0058, -20, 20.8)$.
}
    
    \label{fig_3}
\end{figure}
\begin{figure*}
	\includegraphics[width=18 cm]{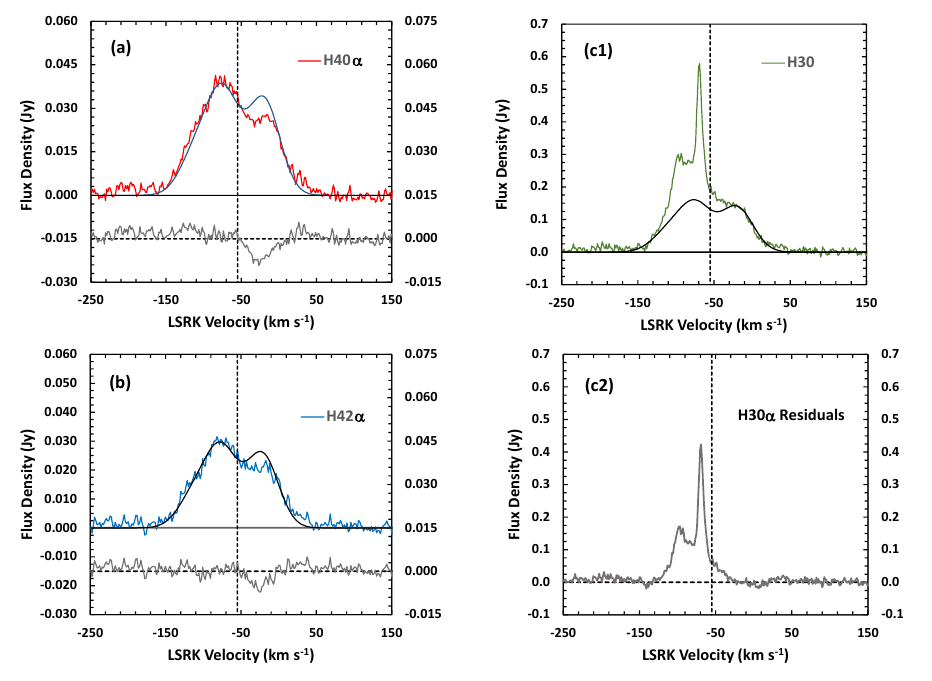}
    \caption{(a) upper graph shows the observed H40$\alpha$ line of Mz3, superposed to a model line, constructed assuming that it was formed in LTE from the H50$\beta$ model profile presented in Fig. \ref{fig_3}. The flux density scale is presented in the left vertical axis; lower graph: difference between observed and model profiles, showing the absorption due to the dasar effect. The flux density scale is presented in the right vertical axis. (b) the same as (a) for the H42$\alpha$ line; (c1)  observed H30$\alpha$ line, superposed to the LTE model line; (c2) difference between observed and model H30$\alpha$ line profiles, where the maser lines can be seen. }
    \label{fig_4}
\end{figure*}

\begin{figure*}
\centering
	\includegraphics[width=17 cm]{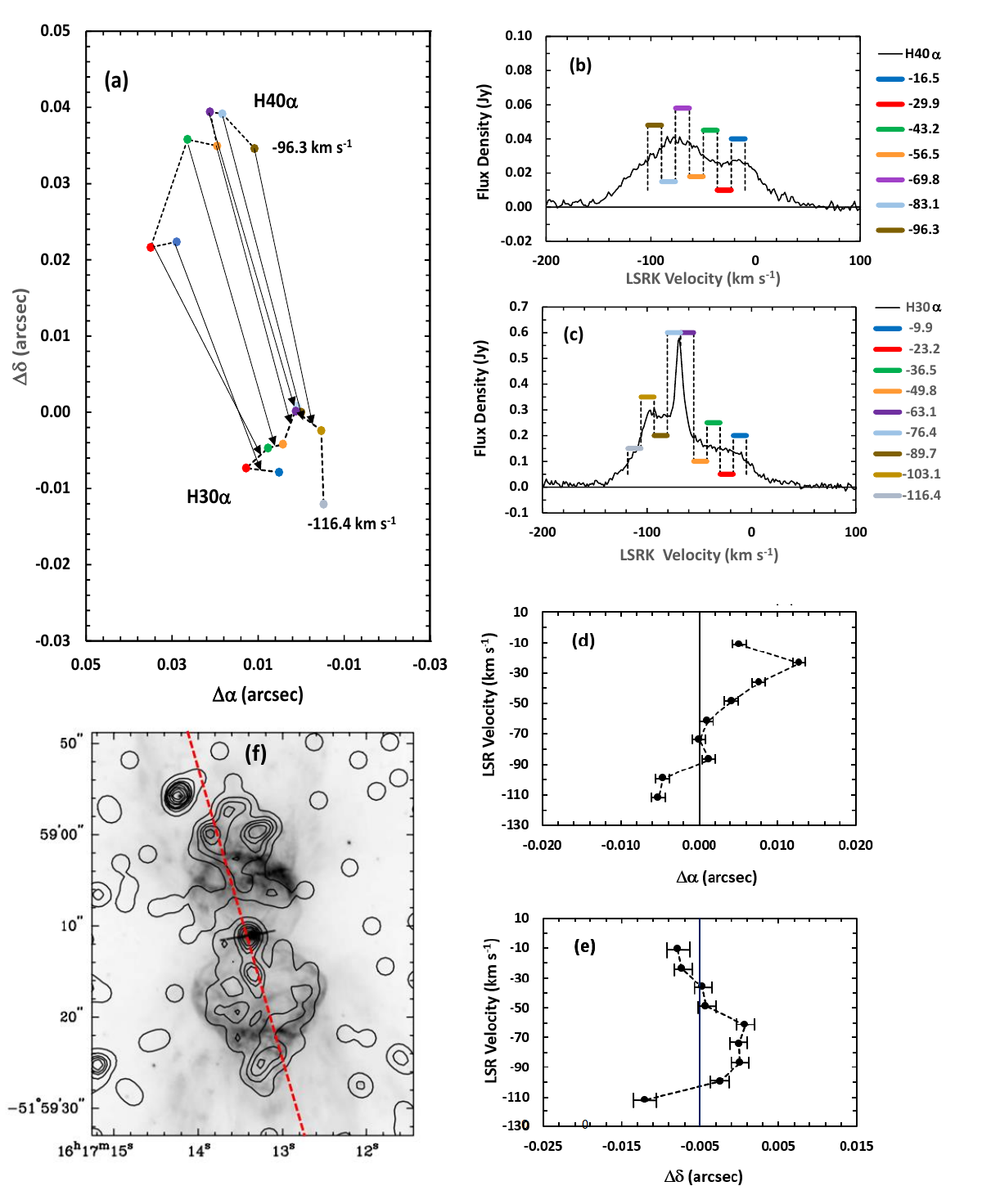}
\textbf{}    \caption{Panel (a): positions of the centroids of a double gaussian model for each of the velocity binned  images of the H40$\alpha$ and H30$\alpha$ lines. The axis represent offsets from  RA $16^{\rm h} 17^{\rm m},   13.381^{\rm s}$, DEC $-51^{\circ} 59' 10\farcs802$, The different colors correspond to different velocities. The  relative errors of the coordinates are always smaller than 1.5 mas, and are shown in panels (d) and (e). The errors in the absolute astrometric coordinates are of the order of $20$ mas.} The black arrows join points of approximately the same velocity in both lines. Panels (b) and (c): profiles of the H40$\alpha$ and H30$\alpha$ lines, with the colored horizontal lines representing the velocity intervals used to determine the positions shown in (a).  Panels (d) and (e): LSR velocities vs RA and DEC for the H30$\alpha$ line. Vertical lines represent the position of the centroid of the continuum image. Panel (f): optical and X-ray image of Mz3 \citep{Guerrero_etal_2004} 
   , with a broken line showing the direction  perpendicular to the linear source structure.
    \label{fig_5}
\end{figure*}

\begin{figure}
\includegraphics[width=8.5 cm]{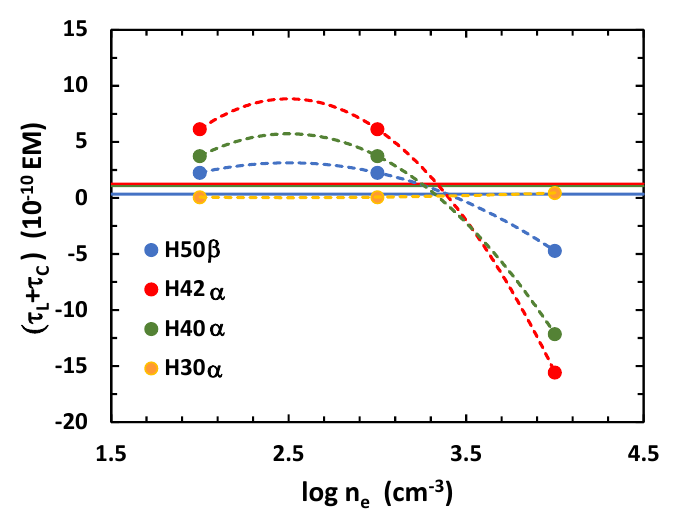}
\textbf{}    \caption Optical depth of the H30$\alpha$, H40$\alpha$, H42$\alpha$ and H50$\beta$ lines for an electron temperature of $7500$ K and different values of the electron density (circles). Dashed lines are quadratic fittings to the data. Horizontal lines represent the optical depth values assuming LTE.
 \label{fig_6}
\end{figure}
The mid infrared emission from its core is consistent with the presence of an edge-on circumstellar disc with its axis inclined by  $ \sim 74^{\circ} \pm   3^{\circ}$ in relation to the line of sight, as determined by \citet{Chesneau_etal_2007}. The disc obscures the central stellar system from our view \citep{Cohen_etal_1978,Chesneau_etal_2007}.

The symmetrical bipolar structure may be considered an indirect evidence of a binary stellar system in its core. This object has been classified as a young planetary nebula (PN), a pre-PN, a symbiotic star and a Symbiotic Mira \citep[]{Cohen_etal_1978,1983MNRAS.204..203L,2001A&A...377L..18S, 2004MNRAS.354..549B, Guerrero_etal_2004, 2019ApJS..240...21A}, but its true nature is still largely unknown.
The parallax of the central star of Mz3 was measured by GAIA Space Observatory as $0.3658 \pm 0.0856$ mas \citep{gai22} corresponding to a distance of $2.73 +0.84/-0.52)$ kpc \citep{bai21}. 

The parameters of the ionizing source have been inferred from the outflow emission. The ionization of the lobes can be explained by a source with effective temperature $T_\textrm{eff} =$~30-40~kK and luminosity of $L_\star \sim 10^4  ~$L$_\odot$ \citep{2003MNRAS.342..383S, 2005A&A...444..861P}. This value may be  underestimated as the dense material in the core may be obscuring the ionizing source. While the lobes have densities around $\sim$4$\times$10$^3$~cm$^{-3}$, the density in the core exceeds 10$^6$~cm$^{-3}$ \citep{2002MNRAS.337..499Z, 2003MNRAS.342..383S}.
\citet{2003ApJ...591L..37K} discovered diffuse X-ray emission within the lobes (collimated outflows, perhaps a jet-like structure) and a compact X-ray source in the core of Mz3. The central source can be an intrinsically hard X-ray source or deeply embedded in X-ray–absorbing gas. Those authors suggested that this emission configuration is consistent with an accretion disk.
The far infrared to submillimeter spectrum (55 to 605~$\mu$m) was obtained with the Herschel Space Observatory as part of the Herschel Planetary Nebula Survey \citep[HerPlaNS;][]{2014A&A...565A..36U}. In addition to the usual bright ionic forbidden lines typically seen in ionized nebulae in this range, its spectrum exhibits unusually bright hydrogen recombination lines (HRLs). \citet{2014A&A...566A..79A} proposed that stimulated emission amplification (laser effect) produced in the dense ionized gas in the core of Mz3 could explain the high intensity of those lines.

The range of physical conditions where stimulated emission amplification can be significant is limited \citep[e.g.,][]{1996ApJ...470.1118S,2000A&A...361.1169H}. 
HRL laser emission can be produced in dense and compact ionized regions (densities higher than 10$^6$~cm$^{-3}$), making of them an interesting tool for inferring physical conditions and kinematics where the forbidden lines typically used as diagnostics are suppressed by the significant collision excitation process due to the densities higher than the critical densities; 
they are so far a rare detection. Only a few objects show this emission:  MWC~349 (B[e] SG star), first RRL maser detected by \citet{1989A&A...215L..13M}, MWC~922 (B[e] FS CMa star), $\eta$~Carinae (luminous blue variable), Cepheus~A~HW2 (young stellar object), MonR2-IRS2 (young stellar object), G45.47+0.05 (young stellar object), and M~82 (starburst galaxy) \citep{ 1995A&A...295L..39C,  1996ApJ...465..691S,   2011ApJ...732L..27J, 2013ApJ...764L...4J,2017A&A...603A..67S, Zhang_etal_2019maser,abr14,abr20}. 


We have carried out ALMA observations of HRLs in Mz3 in order to understand the properties of its central ionized region. We have discovered evidence of maser and, unexpectedly, dasar emission (darkness amplification by stimulated emission) from this region. This is the first time that dasar emission, theoretically predicted almost 3 decades ago by \citet{1996ApJ...470.1118S}, has been identified in any astronomical object. The simultaneous detection of maser and dasar emission  is important for the accurate determination of the densities in both Galactic and extragalatic sources such as compact HII regions, the winds of Wolf-Rayet and Be stars, Active Galactic Nuclei and Starburst Galaxies \citep{1996ApJ...470.1118S}.
The observations are presented in Section \ref{observations}, results are described in Section \ref{results} and discussed in Section \ref{discussion}. Conclusions are summarised in Section \ref{conclusions}.

\section{Observations and data analysis}
\label{observations}
Menzel~3\footnote{ICRS coordinates: R.A. $16^{\rm h} 17^{\rm m} 13.391^{\rm s}$, DEC $-51^{\circ}$ 59$\arcmin$ $10\farcs712$,  proper motions $-3.549 \pm 0.100$ mas y$^{-1}$ in RA and $-3.914 \pm 0.082$ mas y$^{-1}$ in DEC \citep{gai22}.} was observed with the ALMA 12~m array, in bands 3 and 6, during Cycle 8 (project code 2021.1.00912.S). 
 
  The band 3 observations were performed on   2021 November 5-7, totaling  74 minutes on both blocks,  using four 1 GHz SPWs with 1920 channels, 488 kHz  each. The central frequencies were 86.156, 87.741, 98.557 and 100.034 GHz.  
The source J1617-5848 was used as a bandpass and flux calibrator, while sources J1604-441 and J1603-4904 were used as complex gain calibrators. In this band we detected the H40$\alpha$, H42$\alpha$, H50$\beta$ and H57$\gamma$ lines and the underlying continuum.

 The band 6 data were taken on  2022 August 27, using four 2 GHz SPWs, three with 128 channels, centered on 217.991, 218.773, and 230.902 GHz to detect the continuum emission and one 1920 MHz SPW with  high-velocity resolution and 976 kHz-channels centered at 230.957 GHz.
  
Band 6 data were obtained  using the same bandpass and flux calibrator than Band 3, and J1603-4904 as a phase calibrator. We detected the H30$\alpha$ line and the underlying continuum.
   
In both bands we used the  Hoogbom algorithm in the {\it tclean} task of CASA\footnote{CASA: Common Astronomy Software Applications for Radio Astronomy, \citep{cas22}},  Briggs weighting of 0, and a  {\bf pixel size} of 4 milliarcsec; the channel width  resulted in   1.26 km s$^{-1}$ resolution for H30$\alpha$, 1.48km s$^{-1}$   for H40$\alpha$, H50$\beta$ and H57$\gamma$,  and 1.71 km s$^{-1}$ for H42$\alpha$. The corresponding  synthesized beam sizes are  ($0\farcs396\times 0\farcs343, -79^{\circ}$), $(0\farcs252 \times 0\farcs189, -61^{\circ})$ and $(0\farcs289 \times 0\farcs219, -66^{\circ})$.
  
We also cleaned a cube with 12.6  km s$^{-1}$ resolution (10 channels)  for the SPWs that include  the H30$\alpha$ line, and  13.3 and 15.4 km s$^{-1}$  (9 channels) for the H40$\alpha$  and H42$\alpha$ lines, respectively,  to obtain   relative astrometric positions of the centroid of the image for each velocity channel, with a better signal to noise ratio.

  
\section{Results}
\label{results}

Figure \ref{fig_1} presents the band 3 continuum image, with resolution $(0\farcs225 \times 0\farcs165)$. This was the best resolution obtained in the observations and we can see that the source was not resolved. The integrated flux density is $0.12 \pm 0.01$ Jy at 93 GHz and $0.17 \pm 0.03$ Jy at 224 GHz, resulting in a spectral index of 
$0.39^{+0.26}_{-0.31}$. 
 This value is compatible with the emission of an ionized wind or of a compact HII region with the turnover of the bremsstrahlung spectrum above 90 GHz, although there is still  the possibility  that the continuum spectrum is  the combination of the spectra of more than one component, as discussed in Section \ref{other_sources}.

Figure \ref{fig_2} presents the spectra of the detected lines (H30$\alpha$, H40$\alpha$, H42$\alpha$, H50$\beta$ and  H57$\gamma$). The vertical lines show the frequencies of the corresponding H and He recombination lines.  The corresponding frequencies in the LSR reference frame differ by a factor  $7\times 10^{-6}$ of the rest frequency, corresponding to the velocity of  about 2 km s$^{-1}$ of the LSR relative to the baricenter.
The presence of maser effect in the H30$\alpha$ recombination line becomes  clear when we compare its profile with those of the H40$\alpha$ and H42$\alpha$ lines.
Besides, we must point out the difference in the shapes of the H50$\beta$  and H40$\alpha$ lines, the former having a flatter profile.

 A very weak feature  can be seen at the high frequency end of the H40$\alpha$ and H42$\alpha$ lines and could be part of the corresponding He lines; if the half power width of the H and He recombination lines is the same, part of the He line would be blended with the corresponding H line. 
 The helium lines are remarkable faint compared to
    the published helium abundance \citep[$1/5^{\rm th}$ of H,][]{2005A&A...444..861P}, probably reflecting the ionization condition of the region, but  they are too weak to be analysed in detail. 

In order to obtain a quantitative evaluation of the magnitude of the maser effect in the different recombination lines, we use the fact that when the physical conditions  are appropriate for maser amplification in the H30$\alpha$ line, the  H50$\beta$ line is formed in LTE \citep{1996ApJ...470.1118S}. 



To calculate quantitatively the amplification of the H$n\alpha$ lines, with $n= 30, 40$ and 42, we can 
use the observed H50$\beta$  line to estimate the profile that each of the H$n\alpha$ lines would have if they were emitted  in LTE. Since the H50$\beta$ line is too noisy, we fitted three gaussians to better describe its profile.  
 Figure \ref{fig_3} shows the observed H50$\beta$ line profile, in  the LSR velocity scale, the  gaussians,  the model profile (sum of the  gaussians) and the residuals of the fitting. The vertical line shows the systemic LSR velocity of the nebula, $-53.4$ km s$^{-1}$ ($-55$ km s$^{-1}$  heliocentric), as determined by \citet{red00}. 

We  used the model profile $S_{50\beta}^{\rm model}(\nu)$ of the observed H50$\beta$ line to calculate the expected line profiles in LTE  of the H${n\alpha}$ lines, $S_{n\alpha}^{\rm LTE}(\nu)$, using  the following expression: 
\begin{equation}
    S_{n\alpha}^{\rm LTE}(\nu)={C(n\alpha,50\beta}) S_{50\beta}^{\rm model}(\nu)
    \label{eq_1}
\end{equation}
\noindent
with
\begin{equation}
C(n\alpha, 50\beta)=\frac{f_{n+1,n}}{f_{52,50}}\bigg(\frac{52}{50}\bigg)^2\bigg(\frac{n}{n+1}\bigg)^2\bigg(\frac{\nu_{n\alpha}}{\nu_{50\beta}}\bigg)^2,
\end{equation}
\noindent
where $f_{m,n}$ is the oscillator strength of the transition between levels $n$ and $m$ and $\nu_n$ the transition frequency. The values of $C( n\alpha, 50\beta)$ are 23.8, 5.3, and 4.6 for the H30$\alpha$, H40$\alpha$ and H42$\alpha$ lines, respectively \citep{gol68}.

Figure \ref{fig_4} shows the difference between the H$n\alpha$ observed and  model  LTE profiles, obtained from equation \ref{eq_1}, and the respective  residuals. 
The strong narrow lines in the H30$\alpha$ residuals confirm the maser nature of the emission. However,  the negative residuals seen in the H40$\alpha$ and H42$\alpha$ lines, centered at LSR velocity $-25$ km s$^{-1}$, were unexpected; they imply that the intensities of these lines are smaller than those of the lines in LTE. This condition, defined as overcooling or dasar effect, was introduced for the first time  by \citet{1996ApJ...470.1118S} in the theory of HRLs; it  appeared  when  the  angular momentum $l$  was taken into account in the  calculation of $b_ n$, the population of the levels of quantum number $n$ relative to their LTE values  \citep{sto95}. The dasar effect implies that the excitation temperature $T_x$ of the line is  lower than the kinetic temperature, while the maser effect results in negative excitation temperatures, which are defined as:
\begin{equation}
    T_x=\frac{h\nu_0/k}{\ln(n_1/n_2)},   
\end{equation}
 \noindent
 where $\nu_0$ is the frequency of the line, $n_i$ the population density per one degeneracy sublevel, and subindices 1 and 2 denote the lower and upper levels of the transition.
 The ratio $n_1/n_2$ can be written in terms of their LTE values as:
 \begin{equation}
     \frac{n_1}{n_2}=\frac{b_1}{b_2}\frac{n^*_1}{n^*_2},
 \end{equation}
 \noindent
 where * indicates LTE values \citep{1996ApJ...470.1118S}.
 A possible model that explains the simultaneous existence of masers and dasars in the HRLs is presented in Section \ref{masar-dasar}.

\section{Discussion}
\label{discussion}
 Mz3 was not resolved by ALMA, either in the continuum of 93 and 224 GHz, or in the HRLs H30$\alpha$, H40$\alpha$, H42$\alpha$ or H50$\beta$. Interferometric observations of the  8.64 GHz continuum, with resolution $(1\farcs2 \times 0\farcs9)$ obtained on 2002 December 13-15, with the Australian Compact Array, ATCA, \citep{2004MNRAS.354..549B}  also revealed a compact unresolved source with  flux density of  0.20 mJy  and spectral index  between 4.8 and 8.6 GHz of 0.36, obtained from Gaussian fitting to the uniformly weighted radio maps. Assuming that the compact emission detected by ATCA and ALMA are produced by the same source, we obtain a  spectral index between  8.6   and 93 GHz of 0.45, compatible with the value of $0.39^{+0.26}_{-0.31}$ for frequencies between 93 and 224 GHz, presented in Section \ref{results}. 

From the flux density at 93 GHz and assuming the optical depth of the continuum as $\tau_{\rm C} = 1$ and the electron temperature $T_{\rm e}=10^4$ K, we found the solid angle of the source as $\Omega = 7.5 \times 10^{-14}$ sr, which corresponds to an angular radius of $0\farcs03$  or a linear radius $R=80$ a.u. at a distance of 2.7 kpc. Assuming  this value for the length of the source in the direction of the observer, we find  an electron density  $n_{\rm e} = 10^7$ cm$^{-3}$, compatible with the existence of high excitation lines of heavy elements, as detected by \citet{Zhang_etal_2019maser} and with what is expected from the existence of maser effects in the HRLs. 
\subsection{The structure of the emitting region}
\label{structure}

The detection of maser and dasar effects in the HRLs of Mz3  put constrains on the physical conditions of the emitting regions. The difference in the LSR velocities of the emission and absorption residuals implies that they can originate in different sources. In fact, according to \citet{1996ApJ...470.1118S} masers in the H30$\alpha$ line occur at densities of the order of 10$^7$ cm$^{-3}$, while dasars occur in the H40$\alpha$ and H42$\alpha$ lines at densities of the order of 10$^3$ cm$^{-3}$. 
The LSR velocities of the maser lines are  -69 and -98 km s$^{-1}$, and -25 km s$^{-1}$ for the dasar, the three of them close to the velocities used for the  gaussians in the H50$\beta$ line model.  The  velocities, relative to the systemic motion of the nebula, are $-15.6$ and $-44.6$ km s$^{-1}$ for the masers, and   $+28.4$ km s$^{-1}$ for the dasar.

To investigate if the lines come from different positions in the nebula, we  resampled our cube in bins of  10 channels for the H30$\alpha$ line and 9 channels for H40$\alpha$, resulting in a velocity resolution of  12.6 and 13.3 km s$^{-1}$, respectively.
In each of those cubes, we fitted a two-dimensional gaussian in the image plane of each bin, and obtained the position of its centroid\footnote{We  used the CASA fitting routine {\it fit}}. 

Figure \ref{fig_5} (a)  presents the position of the centroids for each velocity channel, for lines H40$\alpha$  and H30$\alpha$. 
The velocities at each position are indicated by different colours and represented in Figures \ref{fig_5} (b) and (c) as straight lines superposed on the line profiles of the two recombination lines. 

 Figures \ref{fig_5} (d) and (e),  present the velocity vs. RA(J2000) and DEC(J2000)  of the centroids of the H30$\alpha$ images; the vertical lines represent the position of the continuum source on August 2022.
The errors in RA and DEC obtained from the fitting are also shown in these figures; they
are compatible with the thermal noise \citep{tho17},  given by:
\begin{equation}
\sigma_{\theta}=\frac{1}{2}\frac{\theta_{\rm res}}{\Re_{\rm sn}}\sim 1.2~ {\rm mas},
\end{equation}
 where $\theta_{\rm res}$ is the resolution of the interferometer, and $\Re_{\rm sn}$ is the signal to noise ratio, which in our case was $\Re_{\rm sn} \simeq 100$ for each channel of both lines.  Similar errors were obtained for the centroids of the H40$\alpha$ images.

  The calculated positions can be used only to compare the relative positions of the different velocity channel images. The uncertainty in their absolute positions must be evaluated taking into account the prescription given by ALMA:
 \begin{equation}
\sigma_{\theta}^{\rm abs}=\frac{\theta_{\rm res}/9}{\min(\Re_{\rm sn},20)}\sim 25~ {\rm mas}.  
\end{equation}

 Therefore, the absolute positions of the H30$\alpha$ and H40$\alpha$ lines shown in Fig. \ref{fig_5}(a) coincide within the precision of the ALMA observations.

 Figures \ref{fig_5} (d) and (e) also show the central position of the continuum source, its ICRS coordinates differ from those of Mz3 by $0.005$ s in RA and $0\farcs 004$ in DEC, showing that the radio source is very close to the star.

In Figure \ref{fig_5}(a) we can see an  almost linear structure, extending for about $0\farcs02$, or 54 a.u. at a distance of 2.7 kpc, with a well defined velocity gradient.
The positions corresponding to the velocity bins of the strongest maser line are very close, indicating that the solid angle of the source is very small, as usual in maser sources.
 The position of the dasar source, with LSR velocity around $-25$ km s$^{-1}$ seems  to be separated from that of the masers, but it is not possible to know if there is still a third source, which corresponds to the emission profile seen in the H50$\beta$ line.  In Figure \ref{fig_5} (f) we present the X-ray and optical images of the extended hourglass nebula \citep{{2003ApJ...591L..37K},Guerrero_etal_2004}, indicating the direction perpendicular to this linear structure.
 


 \subsection{Comparison with other sources}
 \label{other_sources}

 There is a close resemblance between the HRLs profiles of Mz3, presented in this work, and those of the B[e]-type stars MWC 349A  and MWC 922. They all present two narrow lines in the H30$\alpha$ spectrum, attributed to maser amplification, as well as a bipolar nebula surrounding the stars \citep{1989A&A...215L..13M,gva12,2017A&A...603A..67S}.
RCW 349A was extensively studied with interferometric techniques, it was resolved with ALMA in the continuum at the frequency of  345 GHz and the position of the maser sources determined by the centroids of the Gaussian images along the profiles of the H26$\alpha$ and H30$\alpha$ RRLs \citep{pras23}. The authors reported a linear structure,  interpreted as an edge on disk, rotating with Keplerian velocities around the central star, and a low velocity wind, photoevaporated from the disk. A similar model was used to fit the ALMA observations  of MWC 922 \citep{2017A&A...603A..67S}.
 The linear structure defined by the centroids of the velocity channel images of Mz3 are also compatible with rotation, but the low spatial resolution does not allow to define the rotation model. Notice that both masers lines are blushifted in Mz3, while one of them is redshifted in MWC349A.

 
 However, the possible existence of several isolated sources in Mz3 cannot be ignored, an example being the strong HRL maser  lines in $\eta$ Carinae \citep{abr20}, which turned out to be the superposition of 16 spatially resolved sources  with similar velocities, in a region of $0\farcs6$.


\subsection{The coexistence of masers and dasars}
\label{masar-dasar}

 The simultaneous observation of masers and dasar absorption in the HRLs of Mz3  can be explained if we assume that the anomalous absorption is produced by an ionized low-density ($\sim10^3$ cm$^{-3})$  cloud, with temperature of about $7\times 10^3$ K, located between the maser and the observer, moving with a radial LSR velocity of  $-25$ km s$^{-1}$ (28 km s$^{-1}$ relative to Mz3).
 The maser source can be either a dense rotating disk or the combination of several compact sources, with electron densities and temperatures  of the order of $10^7$ cm$^{-3}$ and $10^4$ K, respectively. The existence of the low density clouds is  supported by the extended radio continuum detected with ATCA \citep{2004MNRAS.354..549B}, the physical  conditions of the plasma in Mz3 derived from the optical spectrum \citep{Zhang_etal_2019maser},  and the accretion disk postulated by \citet{2003ApJ...591L..37K} to explain the X-ray absorption in the core,  although the redshifted velocity of the dasar could point out to an external origin  for the cloud.

The observed brightness temperature $T_{\rm B}^{\rm obs}$, with the background emission subtracted, will be:
\begin{equation}
\label{eq6}
    \Delta T_{\rm B}^{\rm obs}(\nu)=[T_{\rm B}^{\rm das}(\nu)-T_{\rm B}^{\rm mas}(\nu)][1-e^{-\tau(\nu)}]
\end{equation}
\noindent
 where $T_{\rm B}^{\rm das}$ and $T_{\rm B}^{\rm mas}$ are the brightness temperatures of the dasar and background maser sources, respectively, and $\tau^{\rm das} = \tau_L^{\rm das} + \tau_C^{\rm das}$ represents the optical depth of the dasar (line plus continuum).

Since $T_{\rm B}^{\rm das}(\nu)<T_{\rm B}^{\rm mas}(\nu)$,  we will see the line in absorption, and assuming $\tau^{\rm das} << 1$  equation \ref{eq6}  becomes: 
\begin{equation}
     \Delta T_{\rm B}^{\rm obs}(\nu)=[T_{\rm B}^{\rm das}(\nu)-T_{\rm B}^{\rm mas}(\nu)][\tau_{\rm L}^{\rm das}(\nu)+\tau_{\rm C}^{\rm das}(\nu)]
\end{equation}
\noindent
 with $ \tau_C^{\rm das}$ and  $\tau_L^{\rm das}$ given by \citet{bro71}:
\begin{equation}
\tau_{\rm C}^{\rm das}= 6.94\times 10^{-8}\nu^{-2}\bigg(\frac{10^4}{T_{\rm e}}\bigg)^{3/2}g_{ff}EM
\end{equation}
\begin{equation}
    \tau_{\rm L}^{\rm das}=1.064\times 10^{-12}b_{\rm n}\beta_{\rm n,n+m}\bigg(\frac{10^4}{T_{\rm e}}\bigg)^{5/2}\frac{f_{\rm n,n+m}}{\nu}(\nu \phi_\nu)EM
\end{equation}
\noindent
 where $g_{ff}$ is the Gaunt factor for free-free emission, $EM=n_{\rm e}^2L$ is the emission measure, $\phi_\nu$ the line profile and $\beta_{\rm n, n+m}$ the amplification factor, given by:
\begin{equation}
    \beta_{\rm n,n+m}=\frac{1-b_{\rm n+m}/b_{\rm n}e^{-h\nu_{\rm n}/kT_{\rm e}}}{1-e^{-h\nu_{\rm n}/kT_{\rm e}}};
\end{equation}
\noindent
 $\phi$ is the line profile, which we assume to be a Gaussian, with a half power width $\Delta\nu$, corresponding to a velocity interval $\Delta v$
\begin{equation}
    \nu\phi_\nu=\bigg(\frac{4\ln 2}{\pi}\bigg)^{1/2}\bigg(\frac{\nu}{\Delta\nu}\bigg)=\bigg(\frac{4\ln 2}{\pi}\bigg)^{1/2}\bigg(\frac{c}{\Delta v}\bigg)
\end{equation}

 The values of $\tau_{\rm C}^{\rm das}(\nu)+\tau_{\rm L}^{\rm das}(\nu)$ at the centers of the H30$\alpha$, H40$\alpha$, H42$\alpha$ and H50$\beta$ lines are presented in Figure \ref{fig_6}, for the electron temperature $7500$ K, $\Delta v = 20$ km s$^{-1}$ and electron densities $10^2$, $10^3$ and $10^4$ cm$^{-3}$. They were calculated using the $b_n$ coefficients published by \citet{sto95} and the Gaunt factors provided by \citet{hoo14}. The LTE values of the optical depths are shown as horizontal lines.

We can see that the values of $\tau^{\rm das}$ differ from those in LTE by different factors, depending on the value of $n$, being higher for the H40$\alpha$ and H42$\alpha$ lines than that for  H30$\alpha$ and H50$\beta$, explaining the dasar effect.

Notice that the values of $b_n$ do not include either the continuum field produced by the emitting plasma or the background photon field,  and therefore the values of the optical depths are only approximations \citep{zhu22}.


\section{Conclusions}
The recombination lines H40$\alpha$, H42$\alpha$, H50$\beta$ and H56$\gamma$ where detected with ALMA in Mz3 on November 2021, while the H30$\alpha$ line was detected on August 2022. Comparison of the profiles and line intensities led to the following conclusions:
\begin{itemize}
    \item 
The H30$\alpha$ line shows clear signs of maser amplification, with two strong lines at velocities of $-16$ and $-46$ km s$^{-1}$, relative to the systemic velocity of the nebula. 
\item
The density of the gas responsible for the maser emission must of the order of $10^7$ cm$^{-3}$. At this density, the H50$\beta$ line is formed in LTE.
\item
The line profiles of the H40$\alpha$ and H42$\alpha$ lines are different from that of the H50$\beta$ line, showing lower intensity than expected in LTE at velocities close to $+28$ km s$^{-1}$ relative to the systemic velocity of the nebula, which we interpret as overcooling or dasar effect.
\item
 The coexistence of masers and dasars can be explained if we assume that the absorption occurs in a low density cloud located in front of the maser source.
\item
The density necessary for the occurrence of a dasar is about $10^3$ cm$^{-3}$, implying the existence of at least two different emitting regions.
\item
 Although the source was not spatially resolved, it was possible to model the images of the velocity cube with two dimensional Gaussians and determine their centroids positions.
\item
These positions are distributed along a line with a  well defined velocity gradient, being compatible with rotation.
\item 
 The detection of  dasars in  HRLs, formed in high density regions, shows the importance of including  also other possible lower density regions, with their corresponding velocities, in the models that calculate the line profiles.
\end{itemize}
The determination of the real structure of the recombination line region needs observation with higher spatial resolution,  which can be achieved by ALMA at higher frequencies and more extended configurations.
\label{conclusions}

\section{Acknowledgements}
This paper makes use of the  ALMA data: 2021.1.00912.S, (PI P.P.B.Beaklini). ALMA is a partnership of ESO (representing its member states), NSF (USA) and NINS (Japan),
together with NRC (Canada), MOST and ASIAA (Taiwan), and KASI (Republic of Korea), in
cooperation with the Republic of Chile. The Joint ALMA Observatory is operated by
ESO, AUI/NRAO and NAOJ.
This work has made use of data from the European Space Agency (ESA) mission
{\it Gaia} (\url{https://www.cosmos.esa.int/gaia}), processed by the {\it Gaia}
Data Processing and Analysis Consortium (DPAC,
\url{https://www.cosmos.esa.int/web/gaia/dpac/consortium}). Funding for the DPAC
has been provided by national institutions, in particular the institutions
participating in the {\it Gaia} Multilateral Agreement.
ZA acknowledges Brazilian agencies FAPESP (grant \#2014/07460-0) and CNPq (grant \#304242/2019-5).
AS acknowledges partial support from H.F.R.I (Basic research financing (Horizontal support of all Sciences) under the National Recovery and Resilience Plan \lq\lq Greece 2.0\rq\rq~funded by the European Union) (Project Number: 15665)
I.A. acknowledges support from Coordena\c{c}\~{a}o de Aperfei\c{c}oamento de Pessoal de N\'{i}vel Superior - Brasil (CAPES) -
and from the Program of Academic Research Projects Management, PRPI-USP.
DRG acknowledges partial support from CNPq (313016/2020-8) and  FAPERJ (200.527/2023).
R.S.’s contribution to the research described here was carried out at the Jet Propulsion Laboratory, California Institute of
Technology, under a contract with NASA, and funded in part by NASA via ADAP awards, and multiple HST GO awards from the Space Telescope Science Institute.


\end{document}